\documentclass[twocolumn,superscriptaddress,showpacs,amsfonts]{revtex4-1}
\pdfoutput=1

\usepackage{amsmath}
\usepackage{amssymb}
\usepackage{graphicx}

\begin{document}
\title{Nodal S-wave Superconductivity in Antiferromagnetic Semimetals}
\author{Wojciech Brzezicki}
\affiliation{CNR-SPIN, IT-84084 Fisciano (SA), Italy, and Dipartimento di Fisica
``E. R. Caianiello\textquotedblright, \\
 Universit\'a di Salerno, IT-84084 Fisciano (SA), Italy}
\affiliation{International Research Centre MagTop at Institute of Physics, Polish Academy of Sciences, Aleja Lotników 32/46, PL-02668 Warsaw, Poland}
\author{Mario Cuoco}
\affiliation{CNR-SPIN, IT-84084 Fisciano (SA), Italy, and Dipartimento di Fisica
``E. R. Caianiello\textquotedblright, \\
 Universit\'a di Salerno, IT-84084 Fisciano (SA), Italy}
\date{\today}
\begin{abstract}
We investigate the impact of s-wave spin-singlet pairing on antiferromagnetic semimetals with Dirac points or nodal loops at the Fermi level. The electron pairing is generally shown to convert the semimetal into a tunable nodal superconductor. The changeover from fully gapped to gapless phases is dictated by symmetry properties of the antiferromagnetic-superconducting state that set the occurrence of a large variety of electronic topological transitions. 
We provide a general criterion for predicting a series of transitions between nodal and fully gapped superconducting phases. Different types of antiferromagnetic patterns are then employed to explicitly demonstrate the microscopic mechanisms that
control the character of the quasiparticle spectrum.
These findings unveil a novel type of nodal superconductivity emerging from the interplay of Dirac fermions and conventional forms of ordering.
\end{abstract}

\maketitle
{\it{Introduction.}} After the great impact of topological insulators~\cite{Qi2011,Hasan2010}, there has been a significant expansion towards metals and semimetals (SMs)~\cite{Volovik2009} as well as quantum materials combining topological and conventional forms of order. 
Topological SMs~\cite{Burkov2016} are materials where conduction and valence bands cross in some points or lines in the Brillouin zone and the crossing are protected by certain symmetry of the system and by the presence of ensuing topological invariants. Among them, Dirac SMs are of particular interest, with massless Dirac fermions emerging as low-energy fundamental excitations. 
Due to their intrinsic instability, to guarantee the robustness, symmetry protection is necessary~\cite{Yang2014,CastroNeto,Young2012}, as, for instance, it occurs in graphene.  
The search for new variants of SMs recently highlighted the interplay of Dirac fermions physics and magnetism. 
Indeed, while most of the currently known SMs are non-magnetic, antiferromagnetic (AFM) SMs can be obtained where both time and inversion are broken while their combination is kept~\cite{Tang2016,Brzezicki2017,Wang2017} or due to chiral- ~\cite{Brzezicki2017} and time-symmetry~\cite{Young2017,Brzezicki2017} combined with non-symmorphic transformations. 

Apart from the large variety of fundamental aspects related to Dirac systems, the combination with other type of conventional orders (e.g. magnetism or superconductivity) represents an ideal testbed for achieving new phases of matter and materials for future emergent technologies \cite{Smeikal2017a,Smeikal2017b}. 
For instance, hallmarks of deviations from a conventional behavior arises when s-wave superconductivity is placed in proximity of a Dirac system, as for the helical superconductor generated at the two-dimensional (2D) boundary of a 3D topological insulator, with vortices hosting Majorana fermions(MFs)~\cite{Fu2008} and supersymmetric behavior~\cite{Grover2014,Ponte2014}.
The tantalizing perspective of topological quantum computing based on MFs motivated the proposal of artificial topological superconductors ~\cite{FuKane2008,Sau2010,Lee2009,AliceaReview,BeenakkerReview,FlensbergReview,KotetesClassi,Weng2011} and the observation of MFs signatures in hybrid superconducting devices~\cite{Mourik,Deng,Furdyna,Heiblum,Finck,Churchill,Yazdani,Franke,Pawlak,Albrecht}.
Although non-magnetic Dirac SMs are natural candidates for topological superconductivity~\cite{Aggarwal2016,Xing2016,Oudah2016,Kobayashi2015}, the role of magnetism in this framework is still largely unexplored.
Such observations together with the traditional strong proximity between superconductivity and antiferromagnetism~\cite{Fradkin2015,Demler2004,Sachdev2003,Scalapino2012} in cuprates, iron pnictides, and heavy fermions, pose fundamental questions concerning the impact of pairing on AFM-SMs. Along this direction, the very recent observation of an anomalous coexistence state of antiferromagnetism and superconductivity in monolayer FeTe grown on a topological insulator~\cite{Manna2017}, reinforces the idea that unexpected quantum effects occur in materials platforms that combine Dirac physics, electron pairing and magnetism. 

In this Letter, we unveil the nature of the low energy excitations of an AFM semimetal in the presence of spin-singlet s-wave pairing (SWP). Contrary to the common view of isotropic pairing leading to fully gapped superconductors, we find that an SWP generally converts the AFM-SM into a tunable nodal superconductor (NSC). A series of electronic topological transitions (i.e. Lifshitz type)~\cite{Lifshitz1960} are predicted on the basis of symmetry principles related to the antiferromagnetic-superconducting state. For this purpose, we provide a general criterion that establishes the relation between the excitations spectrum of the AFM-SM and that one of the NSC. 
Then, different types of AFM-SMs are explicitly investigated to demonstrate how to directly {\it shape} the electronic structure of the NSC.

{\it{Chiral symmetries and criterion for nodal excitations.}} We consider a system of multi-orbital 
itinerant electrons that are coupled via an axial spin-symmetric spin-orbital interaction and through magnetic exchange, $J_{H}$, with localized collinear spins in an AFM pattern described by the variable $S_{i}^{z}=\pm1$. Electrons can locally form spin-singlet pairs with amplitude $\Delta_i$ spatially dependent on the position $i$ in the unit cell and compatible with the periodicity of the AFM pattern.  
The Hamiltonian can be then generally expressed as
\begin{eqnarray}
{\cal H}& \equiv &{\cal H}^{\uparrow}+{\cal H}^{\downarrow}+{\cal H}_{\Delta}^{\uparrow\downarrow} \nonumber \\
{\cal H}^{\sigma}&=&-J_{H}\sum_{i}\sigma S_{i}^{z}+H_{t}^{\sigma} \\
{\cal H}_{\Delta}^{\uparrow\downarrow}&=&\sum_{i}\left(\Delta_{i}d_{i,\uparrow}^{\dagger}d_{i,\downarrow}^{\dagger}+\Delta_{i}d_{i,\downarrow}d_{i,\uparrow}\right) \nonumber.
\end{eqnarray}
with ${H}^{t}$ and ${\cal H}_{\Delta}^{\uparrow\downarrow}$ being the kinetic and pairing terms, respectively. The kinetic part can include both spin-independent hoppings and axial symmetric spin-dependent ones \cite{footnote-so} (i.e. spin-orbital coupling as in Refs. \cite{Kane1,Kane2}) in such a way that, for a given AFM pattern with opposite spin domains and an equal number of sites, ${\cal H}^{\sigma}$ has a chiral symmetry, i.e. there exists a unitary operator anticommuting with ${\cal H}^{\sigma}$.   
The chiral symmetry (CS) is a fundamental property for the classification of topological states of matter~\cite{Wen1989,Schnyder2008,Kitaev2009, Koshino2014}. The CS generally relates positive and negative parts of the energy spectrum and it can be associated with the presence of effective sublattices in the phase space. Indeed, one can construct a basis with different eigenvalues of the chiral operator with the Hamiltonian having vanishing matrix elements inside the same sublattice sector. 
\begin{figure}[t]
\includegraphics[clip,width=1\columnwidth]{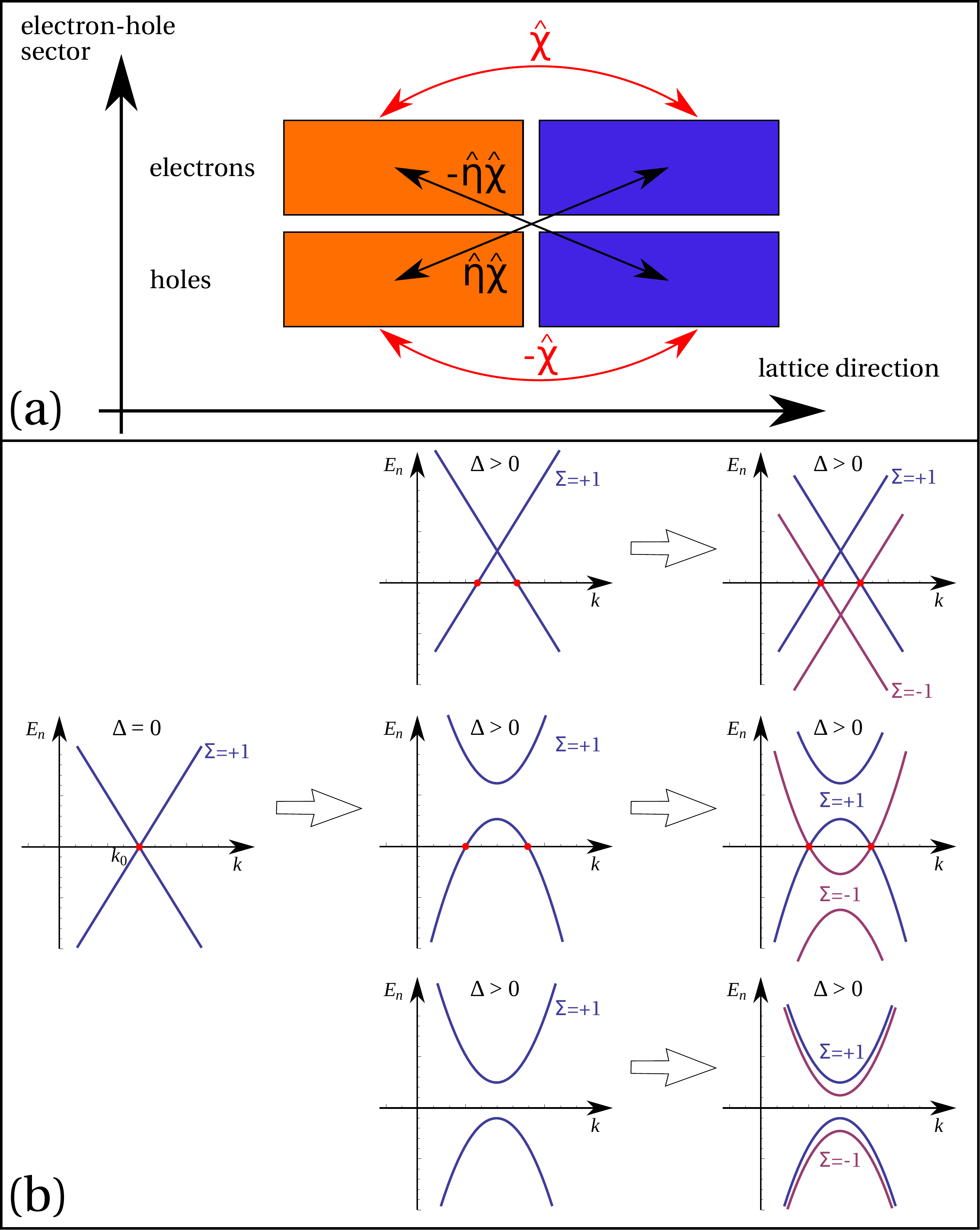}
\protect\caption{(a) Sketch of the transformations involved in the chiral symmetries which link electron or hole and electron-hole sectors in different magnetic domains. (b) Schematic of Dirac points at zero energy and $k=k_{0}$ (red point) in the $\Sigma$ symmetric subsector $\hat{H}_{\vec{k}}^{+}$ 
(left column) and for non-vanishing $\hat{\Delta}$ (middle column).
Different types of gapless or gapfull spectra are shown. 
The right column reports the full
spectrum for both $\Sigma$ sectors, thus restoring the chiral symmetry. \label{fig:zero-band}}
\end{figure}

Hence, the structure of the resulting Hamiltonian in the $k$-space is
\begin{equation}
{\cal H}_{\vec{k}}=\begin{pmatrix}\hat{H}_{\vec{k},\uparrow}^{e} & \hat{\Delta}\\
\hat{\Delta} & \hat{H}_{\vec{k},\downarrow}^{h}
\end{pmatrix},
\end{equation}
where $\hat{H}_{\vec{k},\downarrow}^{h}=-(\hat{H}_{-\vec{k},\downarrow}^{e})^{T}$, while the matrix $\hat{\Delta}$ 
has entries $\Delta_{p}$ ($p$ being the sites label) and a dimension that is set by the number of sites in the unit cell and on-site orbital degrees. 
If ${\cal \hat{\chi}}_{\vec{k}}$
is a CS operator for both particle and hole sectors and 
commutes with $\hat{\Delta}$ then, by construction, we have $\{{\cal H}_{\vec{k}},{\cal S}_{\vec{k}}\}=0$ for ${\cal S}_{\vec{k}}=\hat{\chi}_{\vec{k}} \tau_z$, with $\tau_z$ being the Pauli matrix acting in the particle-hole space.
Due to the AFM structure, ${\cal \hat{\chi}}_{\vec{k}}$ includes a translation of a half-vector of the Bravais lattice, thus it has a non-symmorphic character~\cite{Shiozaki2015}.
Taking into account the structure of the Hamiltonian, another independent CS operator, ${\cal S}'_{\vec{k}}$, can be constructed be introducing a unitary transformation $\hat{\eta}$ with the following properties: i) $[\hat{\eta},\hat{\Delta}$]=0, ii) $\hat{\eta}^{2}=\hat{1}$, and iii) $\hat{\eta}\hat{H}_{\vec{k},\uparrow}^{e}\hat{\eta}=\hat{H}_{\vec{k},\downarrow}^{h}$.
Upon these assumptions ${\cal S}'_{\vec{k}}$ is given by ${\cal S}'_{\vec{k}}=i \hat{\eta} \hat{\chi}_{\vec{k}} \tau_y$.
%
%
\begin{figure*}[t]
\includegraphics[clip,width=1\textwidth]{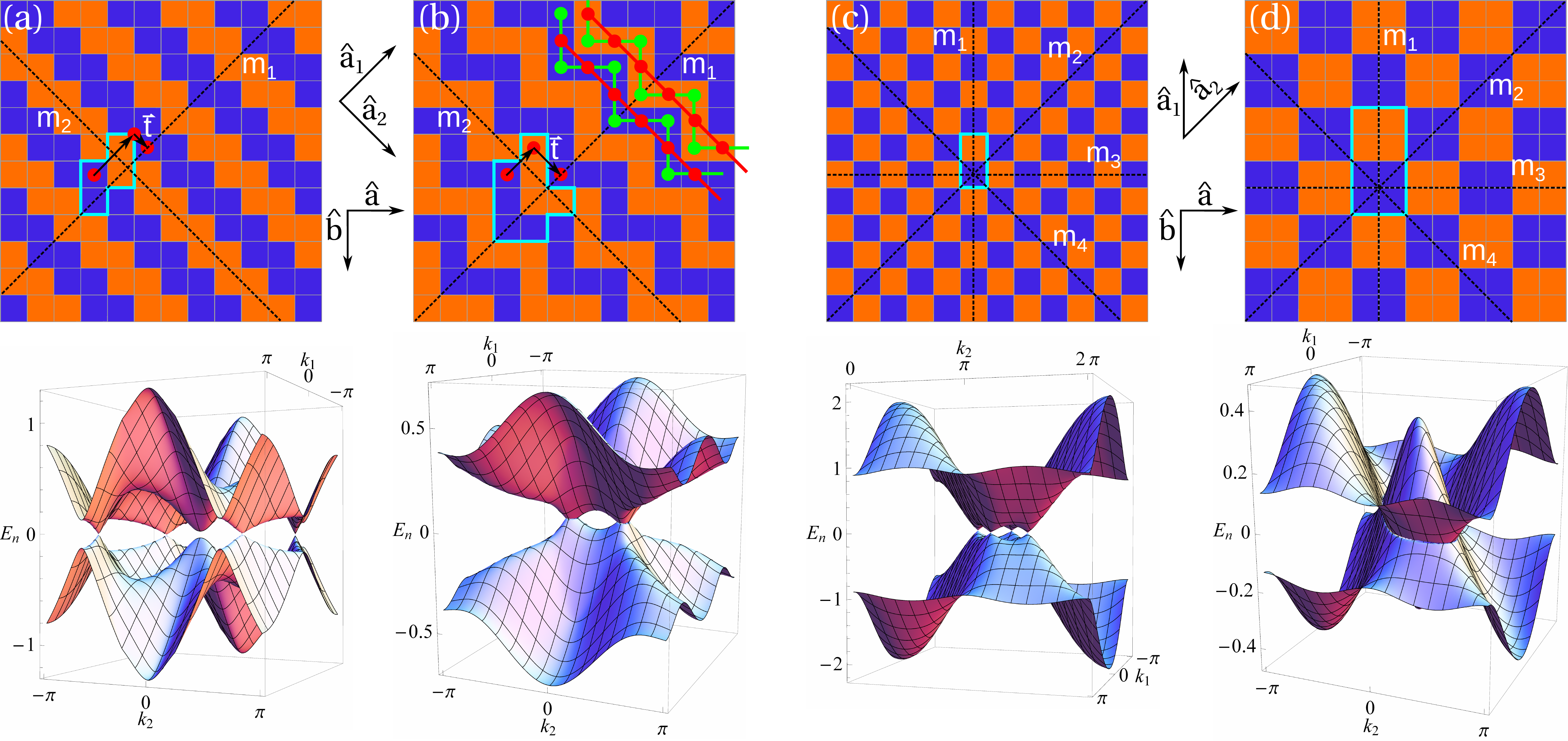}
\protect\caption{
Upper row: schematics of AFM patterns. (a) AFM zig-zag with
length $L_{z}=2$ ($z2$), (b) $L_{z}=3$ ($z3$),
(c) AFM Ne\'el state and (d) checkerboard with $2\times2$ magnetic domains ($c2$).
The orange (blue) squares indicate sites with spin up (down). The unit cell is marked by thick blue
frame and $\hat{a}_{1,2}$ are the basis vectors of the Bravais lattice. Dashed
lines indicate the normal mirror or glide mirror lines when considering the zig-zag AFM patterns ((a) and (b)). 
The gliding transformation is
shown by red dots and arrows, i.e. dot is firstly reflected via
$m_{2}$ and then translated by a vector $\vec{t}$ parallel to
the mirror line $m_{2}$. For $z3$ pattern, red and green lines connect dots related to
the second and third neighbor hopping amplitudes $\pm\delta$
for spin up/down domains, respectively. Bottom row: electronic spectra associated with the corresponding AFM patterns (upper row). For a given spin-polarization 
Dirac points or nodal loop occur at the Fermi level for $\hat{\Delta}=0$.
\label{fig:patt_spec}}
\end{figure*}

In terms of physical transformations, ${\cal S}_{\vec{k}}$ acts to transpose
the magnetic domains for both electrons and holes 
whereas ${\cal S}'_{\vec{k}}$ interchanges
electrons in the $S_{i}^{z}=\pm1$ spin domain with holes in the opposite one, i.e. $S_{i}^{z}=\mp1$ (see Fig 1(a)).
From the properties of $\hat{\eta}$ and ${\cal \hat{\chi}}_{\vec{k}}$
one can observe that they either commute or anticommute, and consequently, a reversed algebra is due for ${\cal S}_{\vec{k}}$ and ${\cal S}'_{\vec{k}}$, respectively. More importantly, since we deal with CS operators, their product $\Sigma={\cal S}_{\vec{k}}{\cal S}'_{\vec{k}}$ leads to a symmetry for the Hamiltonian, i.e. $[\Sigma,{\cal H}_{\vec{k}}]=0$.
Hence, we end up with a unitary operator $\Sigma$ which squares to identity and it has eigenvalues $\pm 1$. 
Remarkably, in the $\Sigma$ eigenbasis the Hamiltonian can be rewritten in such a way that \cite{SupMat} 
\begin{eqnarray}
\hat{H}_{\vec{k}}^{\pm}\equiv\hat{H}_{\vec{k},\downarrow}^{h}\pm\hat{\eta}\hat{\Delta}
\end{eqnarray}
and the pairing term becomes an effective potential that separates in each $\Sigma$ symmetric sector. This is one of the central results of the Letter. In the symmetry projected basis, electrons with a given spin polarization are coupled to the AFM background and interact with an effective potential that arises from the pairing term through the CS derived transformations. 
Interestingly, the induced potential can even break the CS within each $\Sigma$ sector which is, however, recovered when both the spectra of $\hat{H}_{\vec{k}}^{+}$ and $\hat{H}_{\vec{k}}^{-}$ are merged together.  

The structure of the symmetry projected Hamiltonian allows us to deduce important consequences on the evolution of the low-energy excitations spectra. 
Since the AFM-SM has Dirac points (DPs) or line nodes (LNs) at the Fermi level, for such $\vec{k}\in$ points the phase space, at a given spin polarization, is spanned by two vectors $\{|E_{\chi,\vec{k}}^{(0)}\rangle\}$ labeled by the eigenvalues of 
${\cal \hat{\chi}}_{\vec{k}}$, i.e. $\chi=\pm1$. Then, to address the energy correction to the DPs or nodes, one can employ a perturbation approach by expressing the effective pairing interaction $\hat{V}\equiv\hat{\eta}\hat{\Delta}$in the $\vec{k}$-dependent
basis of $\{|E_{\pm,\vec{k}}^{(0)}\rangle\}$ as $\hat{V}_{\chi'\chi,\vec{k}}=\langle E_{\chi',\vec{k}}^{(0)}|\hat{\eta}\hat{\Delta}|E_{\chi,\vec{k}}^{(0)}\rangle \,.$
The eigenvalues of $\hat{V}_{\chi'\chi,\vec{k}}$ matrix provide corrections
to the zero energy Dirac states and, in turn, set their evolution for each $\Sigma$ sector. A schematic view of this process is presented in Fig. 1(b). 
Indeed, except for symmetry protected or accidental cases where $\hat{V}$ eigenvalues are identically vanishing, it is their sign that generally yields the evolution of the DPs at the Fermi level.
With opposite sign eigenvalues, under the action of $\hat{\Delta}$,
the DPs acquire a mass, evolving into positive and negative-energy
states, and thus generally a gap opening occurs (Fig. 1(b)). Otherwise, for eigenvalues having the same sign, the spectrum rigidly shifts above or below zero energy, thus leading to gapless excitations at the Fermi level. In this case, although a gap opening may occur at the Dirac point, a zero-energy crossing of the spectra and gapless modes in the superconducting state are expected (Fig. 1(b)). 
In general we observe that the impact of the pairing on the AFM-SM can be assessed by evaluating the determinant ${\cal D}_{\vec{k}}\equiv\det\hat{V}_{\vec{k}}$ in the whole Brillouin zone. Positive (negative) values for ${\cal D}_{\vec{k}}$ translate into gapless (gapped) modes at a given infinitesimally small amplitude of $\hat{\Delta}$. The sign of ${\cal D}_{\vec{k}}$ thus gives a general criterion for predicting the occurrence of nodal phases and induced Lifshits transitions once the amplitude of $\hat{\Delta}$ is varied. Since there are no symmetry constraints for the evolution of the spectra in the $\Sigma$ symmetric sectors, the emerging SC state can exhibit a large variety of transitions from gapless to gapped configurations within the Brillouin zone. We point out that this result holds independently of the symmetry protection behind the AFM-SM.
\begin{figure*}[t]
\includegraphics[clip,width=1\textwidth]{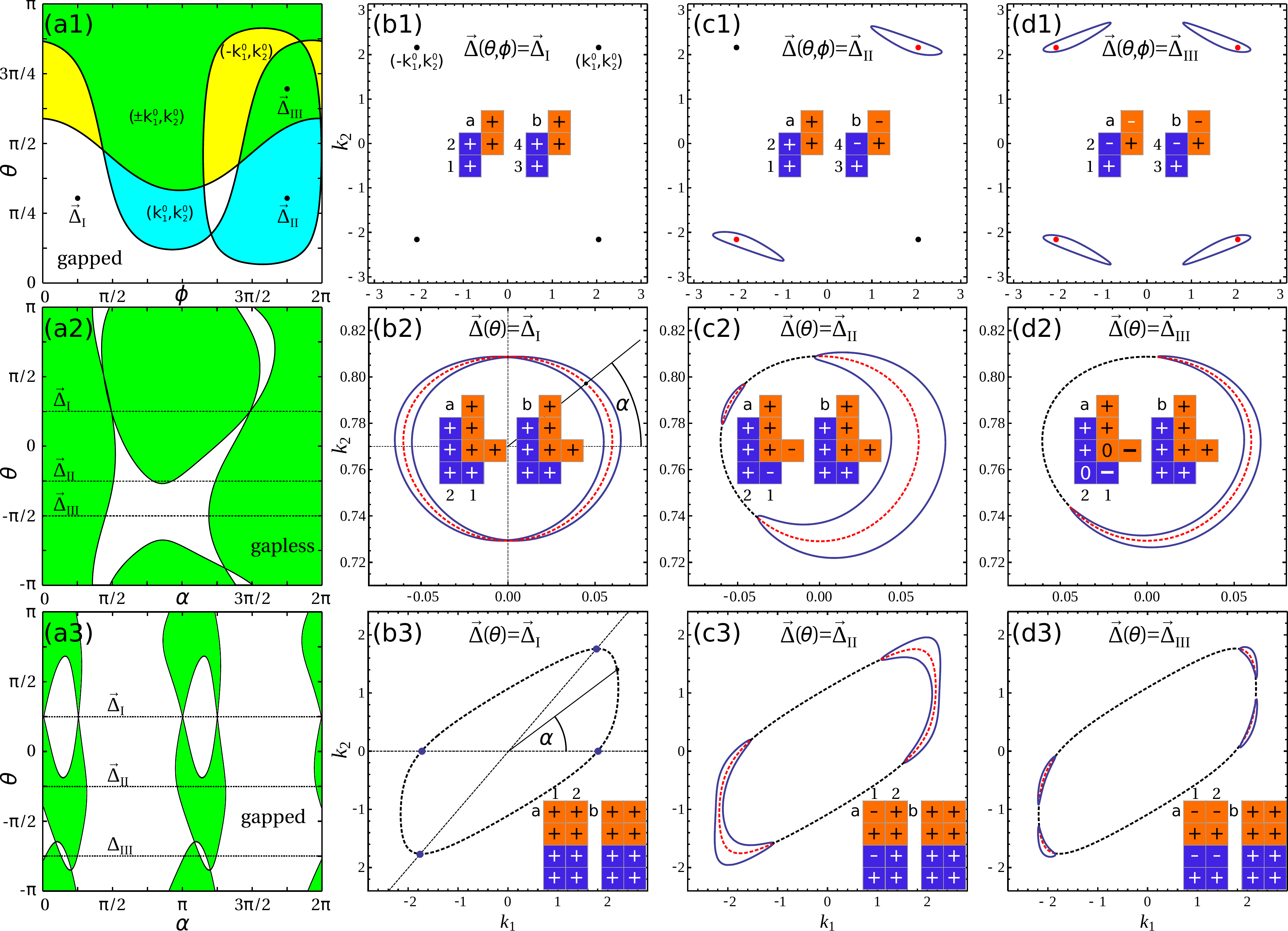}
\protect\caption{
Phase diagrams for gapless (colored) and gapped (white) SC states and induced Lifshitz transitions by varying the pairing amplitude in the unit cell for $z2$ (a1), $z3$ (a2) and $c2$ (a3) AFM patterns. 
Parameterization of the unit cell pairing components ${\Delta}_i$ of ${\vec{\Delta}}$ is made by introducing the angles $\theta$ and $\phi$: $(\Delta_{1},\Delta_{2},\Delta_{3},\Delta_{4})=(1,\sqrt{3}\cos\theta,\sqrt{3}\sin\theta\cos\phi,\sqrt{3}\sin\theta\sin\phi)$ in (a1), $\Delta_{i\neq(1,2)}=1$, $\Delta_{1}=\sqrt{2}\sin\theta$
and $\Delta_{2}=\sqrt{2}\cos\theta$ in (a2) and (a3). 
Representative evolution of the semimetal phase for different configurations of ${\vec{\Delta}}$ in the unit cell (see Figs. 1 (a)-(c)) associated with $\hat{\Delta}_{I,I\! I,I\! I\! I}$
in the phase diagrams (a$i$, $i=1,2,3$): (b1)-(d1) for $z2$, (b2)-(d2) for $z3$, and (b3)-(d3) for $c2$ AFM patterns. 
$\alpha$ indicates the angular position of a given $k$ point at the Fermi level.
Site labels and schematic of the unit cell are in the center of the panels (b$i$),(c$i$),(d$i$) with $i=1,2,3$; $a$ and $b$ refer to the orbitals and ($+$,$-$) indicate the sign of the pairing amplitude. Blue lines or dots in 
(b$i$), (c$i$) and (d$i$) [$i=1,2,3$] denote the nodal superconducting Fermi surface ($\hat{\Delta}\neq 0)$),
black/red lines or dots indicate the Fermi point and the nodal loops for the AFM semimetal ($\hat{\Delta}=0$).
\label{fig:FS3}}
\end{figure*}

{\it{Nodal superconductivity in multi-orbital AFM-SMs.}}
In order to explicitly demonstrate the impact of spin-singlet pairing in chiral symmetric AFM-SMs we employ a model system which allows to have Dirac points or nodal loops at the Fermi level. 
Such states can be realized in an effective 2D orbital-directional double-exchange model \cite{Brzezicki2015,Brzezicki2017}
describing itinerant electrons (e.g. t$_{2g}$ or $p$ bands) in the
presence of an anisotropic spin-orbit coupling, as due to tetragonal crystal field
splitting, and Hund coupled to localized spin moments forming AFM patterns (see Figs. 2(a)-(d)).
The kinetic part $H_{t}^{\sigma}$, as in Eq. 1, is then given by
\begin{equation}
H_{t}^{\sigma}=\sum_{i}\sum_{{\alpha,\beta=a,b\atop {\hat{\gamma}}={\hat{a},\hat{b}}}}\! t_{\gamma,\alpha\beta}\left(d_{i,\alpha\sigma}^{\dagger}d_{i+\hat{\gamma},\beta\sigma}\!+\! h.c.\right)\!+\!\lambda\sum_{i}l_{i}^{z}S_{i}^{z},
\end{equation}
where the only non-vanishing hopping amplitudes are $t_{\hat{a},bb}=t_{\hat{b},aa}=-t$
and $\hat{\gamma}=\hat{a},\hat{b}$ are
bond directions on a square lattice. $\lambda$ is the spin-orbit coupling
in the $(a,b)$ orbitals subspace, with $l_{i}^{z}=i\left(d_{i,a,\sigma}^{\dagger}d_{i,b,\sigma}-d_{i,b,\sigma}^{\dagger}d_{i,a,\sigma}\right)$.
Localized spins $S_{i}^{z}$ can form zig-zag or checkerboard patterns
as depicted in Fig. \ref{fig:patt_spec}. For all the examined AFM configurations, 
the model exhibits two
chiral symmetries with $\hat{\eta}$ being a diagonal matrix with $\pm1$
entries \cite{SupMat}. Since $\hat{\eta}$ commutes
with $\hat{\chi}{}_{\vec{k}}$ \cite{SupMat} when considering zig-zag $z2$, $z3$ or checkerboard $c2$, one can make use of the 
criterion for the sign of ${\cal D}_{\vec{k}}$ to predict the evolution
of the superconducting spectra at any given pairing configuration of $\hat{\Delta}$. On the other hand, for the N\'eel AFM state (Fig. 2(c)), $\hat{\eta}$
anticommutes with $\hat{\chi}{}_{\vec{k}}$. Thus, a finite amplitude of
$\hat{\Delta}$ does not break the chiral symmetry in each $\hat{H}{}_{\vec{k}}^{\pm}$ sector and the criterion does not directly apply. In this case, the SC has point or line nodes in each $\Sigma$ sector \cite{SupMat}.
Since the unit cell in the AFM phases have inequivalent sites, inhomogeneities in the superconducting order parameters are physically relevant for the coexistence state. For this aim, we investigate the role of both amplitude and sign change of the spin-singlet pairing within the various types of AFM-SMs. We find that, as expected from the general analysis, the superconducting state exhibits distinct transitions from gapped to gapless phases (Fig. 3 (a1)-(a3)) that demonstrate a high degree of tunability of the spectra.  
For instance, starting from an AFM-SM with Dirac points (Fig. 3 (b1)) or nodal line (Fig. 3  (b2) or (b3)), a pattern $\hat{\Delta}$ with uniform sign in the unit cell can lead to i) a fully gapped SC (Fig. 3 (b1)), ii) a SC with double nodal Fermi surfaces (Fig. 3 (b2)), iii) or a point node SC (Fig. 3 (b3)). The occurrence of an orbital dependent sign change for the spin-singlet order parameter at one or few sites within the unit cell opens other channels of transitions from a two-nodal (Fig. 3 (c1)) to four-nodal rings developing around the position of the AFM Dirac points (Fig. 3 (d1)). On the other hand, the evolution of the AFM-SM with Dirac nodal loop shows how one can achieve gapless phases with nodal rings that can touch in two points (Fig. 3 (b2)) or are disconnected (Fig. 3 (c2)) and, eventually, exhibits transitions from a two-nodal to one-nodal ring Fermi surface (Fig. 3 (d2). Such phenomenology generally holds also for the $c2$ AFM (Figs. 3 (a3)-(d3)) where the modification of spatial pairing amplitude leads to a series of Lifshitz transitions from point nodes to two- and four-nodal loops Fermi surface (Fig. 3 (d3)). 
%

{\it{Concluding remarks and materials outlook.}} We demonstrate that a completely novel type of NSC can arise by combining an AFM-SM and spin-singlet SWP. 
Concerning the materials cases, in the half-Heusler family superconductivity can coexist with antiferromagnetism \cite{Pan2013, Nakajima2015,Pavlosiuk2016} and SM phases have been predicted \cite{Yu2017}. Moreover, heterostructures made of an AFM-SM interfaced with conventional spin-singlet SC is well suited for observing such form of nodal superconductivity. Although CS in our model holds at a given electron filling in the AFM-SM, for materials purposes CS can generally be preserved in the low-energy sector where superconducting correlations are relevant for setting the coexistence state. Apart from direct signatures of NSC by angle resolved spectroscopy and scanning tunneling microscopy \cite{Qi2011,Hasan2010,Yan2017}, the sensibility to spatial variation of the superconducting order parameter can lead to unique thermodynamical features with a cascade of Lifshitz transitions that get thermally activated. 

{\bf{Acknowledgements}} W.B. acknowledges support by the European Horizon 2020 research and innovation programme under the Marie-Sklodowska Curie grant agreement No. 655515 and support by Narodowe Centrum Nauki (NCN, National Science Centre, Poland) Project No.~2016/23/B/ST3/00839. The research was partially supported by the Foundation for Polish Science through the IRA Programme co-financed by EU within SG OP.


\section{Appendix}

In the Appendix we provide details for the derivation of the criterion for nodal superconductors. For $z2$ spin pattern we explicitly report the matrix structure of $\hat{\eta}$ and $\hat{\chi}$ that build up the chiral symmetry operators. Then, a representative case of the evolution of the superconducting state for the N\'eel AFM state is also reported.

\subsection{Details on zig-zag $2$ system}

For a zig-zag $2$ spin pattern the electron Hamiltonians for the fixed spin channel $\sigma$ have a form of $8\times 8$ matrices given by,
\begin{figure*}
\begin{equation}
\hat{H}_{\vec{k},\uparrow}^{e}=\begin{pmatrix}\sigma J_{H} & -te^{-ik_{2}} & 0 & -t & i\sigma\lambda & 0 & 0 & 0\\
-te^{ik_{2}} & \sigma J_{H} & -te^{-ik_{1}} & 0 & 0 & i\sigma\lambda & 0 & 0\\
0 & -te^{ik_{1}} & -\sigma J_{H} & -te^{-ik_{2}} & 0 & 0 & -i\sigma\lambda & 0\\
-t & 0 & -te^{ik_{2}} & -\sigma J_{H} & 0 & 0 & 0 & -i\sigma\lambda\\
-i\sigma\lambda & 0 & 0 & 0 & \sigma J_{H} & -t & 0 & -e^{-ik_{2}}\\
0 & -i\sigma\lambda & 0 & 0 & -t & \sigma J_{H} & -te^{-i(k_{1}-k_{2})} & 0\\
0 & 0 & i\sigma\lambda & 0 & 0 & -te^{i(k_{1}-k_{2})} & -\sigma J_{H} & -t\\
0 & 0 & 0 & i\sigma\lambda & -e^{ik_{2}} & 0 & -t & -\sigma J_{H}.
\end{pmatrix}
\end{equation}
\end{figure*}
The internal chirality (not related to superconductivity) $\hat{\chi}_{\vec{k}}$ has a form of,
\begin{equation}
\hat{\chi}_{\vec{k}}=\begin{pmatrix}0 & 0 & -e^{-i\frac{k_{1}}{2}} & 0 & 0 & 0 & 0 & 0\\
0 & 0 & 0 & e^{-i\frac{k_{1}}{2}} & 0 & 0 & 0 & 0\\
-e^{i\frac{k_{1}}{2}} & 0 & 0 & 0 & 0 & 0 & 0 & 0\\
0 & e^{i\frac{k_{1}}{2}} & 0 & 0 & 0 & 0 & 0 & 0\\
0 & 0 & 0 & 0 & 0 & 0 & -e^{-i\frac{k_{1}}{2}} & 0\\
0 & 0 & 0 & 0 & 0 & 0 & 0 & e^{-i\frac{k_{1}}{2}}\\
0 & 0 & 0 & 0 & -e^{i\frac{k_{1}}{2}} & 0 & 0 & 0\\
0 & 0 & 0 & 0 & 0 & e^{i\frac{k_{1}}{2}} & 0 & 0,
\end{pmatrix}
\end{equation}
and $\hat{\eta}$ operator connecting the Hamiltonians for electron and hole sectors as $\hat{\eta}\hat{H}_{\vec{k},\uparrow}^{e}\hat{\eta}=\hat{H}_{\vec{k},\downarrow}^{h}$ is, 

\begin{equation}
\hat{\eta}=\begin{pmatrix}-1 & 0 & 0 & 0 & 0 & 0 & 0 & 0\\
0 & 1 & 0 & 0 & 0 & 0 & 0 & 0\\
0 & 0 & -1 & 0 & 0 & 0 & 0 & 0\\
0 & 0 & 0 & 1 & 0 & 0 & 0 & 0\\
0 & 0 & 0 & 0 & 1 & 0 & 0 & 0\\
0 & 0 & 0 & 0 & 0 & -1 & 0 & 0\\
0 & 0 & 0 & 0 & 0 & 0 & 1 & 0\\
0 & 0 & 0 & 0 & 0 & 0 & 0 & -1.
\end{pmatrix}
\end{equation}

\subsection{Criterion for occurance of nodal superconductivity}

We start by demonstrating how to get the effective pairing potential in the $\Sigma$ symmetric representation.
The eigenbasis of $\Sigma$ can be expressed by an orthogonal matrix ${\cal V}_{\Sigma}$
\begin{equation}
{\cal V}_{\Sigma}=\frac{1}{\sqrt{2}}\left(\begin{array}{cc}
-\hat{\eta} & \hat{\eta}\\
\hat{1} & \hat{1}
\end{array}\right),\quad{\cal V}_{\Sigma}^{T}\Sigma{\cal V}_{\Sigma}=\left(\begin{array}{cc}
-\hat{1} & 0\\
0 & \hat{1}
\end{array}\right).
\end{equation}
By means of ${\cal V_{\Sigma}}$ one can rewrite ${\cal H}_{\vec{k}}$ in
a block-diagonal form. Indeed, since $\hat{\eta}^{2}=\hat{1}$ and $\hat{\eta}\hat{H}_{\vec{k},\uparrow}^{e}\hat{\eta}=\hat{H}_{\vec{k},\downarrow}^{h}$, one obtains
\begin{equation}
\bar{{\cal H}}_{\vec{k}}={\cal V}_{\Sigma}^{T}{\cal H}{\cal V}_{\Sigma}=\begin{pmatrix}\hat{H}_{\vec{k},\downarrow}^{h}-\hat{\eta}\hat{\Delta} & 0\\
0 & \hat{H}_{\vec{k},\downarrow}^{h}+\hat{\eta}\hat{\Delta}
\end{pmatrix}.\label{eq:Hambl}
\end{equation}
Thus, the pairing term effectively separates in the $\Sigma$ symmetric sectors. 
When $\hat{\eta}$ commutes with ${\cal \hat{\chi}}_{\vec{k}}$, the action of $\hat{\Delta}$
is to break the chiral symmetry within each sector, $\hat{H}_{\vec{k}}^{\pm}\equiv\hat{H}_{\vec{k},\downarrow}^{h}\pm\hat{\eta}\hat{\Delta}$
of $\bar{{\cal H}}_{\vec{k}}$. 
The overall chiral symmetry of $\bar{{\cal H}}_{\vec{k}}$
is recovered by taking the two $\Sigma$ symmetric blocks together. This means that the spectrum
of $\hat{H}_{\vec{k}}^{+}$ is linked to that of $\hat{H}_{\vec{k}}^{-}$ by a sign reversal.
On the other hand, for $\hat{\eta}$ anticommuting with ${\cal \hat{\chi}}_{\vec{k}}$
both sectors of $\bar{{\cal H}}_{\vec{k}}$ are chiral symmetric and
their spectra are unrelated.

Concerning the derivation of the criterion for establishing the character of the excitations spectra in the superconducting state, we observe that the considered AFM semimetal states exhibit isolated Dirac points or Dirac nodal lines. Then, for any given $k$ at the Fermi level we have double-degenerate configurations expressed as
$\{|E_{\chi,\vec{k}}^{(0)}\rangle\}$, with $\chi=\pm1$ labeling the eigenvalues of the chiral symmetry operator
${\cal \hat{\chi}}_{\vec{k}}$. In the eigenbasis of ${\cal \hat{\chi}}_{\vec{k}}$ the 
Hamiltonian at zero pairing amplitude has a block off-diagonal form 
\begin{equation}
\hat{H}_{\vec{k}}^{\pm}(\hat{\Delta}=0)=\left(\begin{array}{cc}
0 & u_{\vec{k}}\\
u_{\vec{k}}^{\dagger} & 0
\end{array}\right).
\end{equation}
Hence, the states $\{|E_{\chi,\vec{k}}^{(0)}\rangle\}$ are vectors
of the null space for $\hat{H}_{\vec{k}}^{\pm}(\Delta=0)$ and within such basis representation they take
a block form given by 
\begin{equation}
|E_{+,\vec{k}}^{(0)}\rangle=\left(\begin{array}{c}
0\\
|u_{R,\vec{k}}^{(0)}\rangle
\end{array}\right),\quad|E_{-,\vec{k}}^{(0)}\rangle=\left(\begin{array}{c}
|u_{L,\vec{k}}^{(0)}\rangle\\
0
\end{array}\right),\label{eq:zero_states}
\end{equation}
where $|u_{R/L,\vec{k}}^{(0)}\rangle$ are right/left zero states
of $u_{\vec{k}}$, i.e., $u_{\vec{k}}|u_{R,\vec{k}}^{(0)}\rangle=\langle u_{L,\vec{k}}^{(0)}|u_{\vec{k}}=0$.
In the presence of $\hat{\Delta}$, starting from the AFM-SM the lowest-order correction can be obtained by evaluating
$\hat{V}\equiv\hat{\eta}\hat{\Delta}$ in the $\vec{k}$-dependent
basis of $\{|E_{\pm,\vec{k}}^{(0)}\rangle\}$. The matrix elements of the effective pairing potential 
are given by
\begin{equation}
\hat{V}_{\chi'\chi,\vec{k}}=\langle E_{\chi',\vec{k}}^{(0)}|\hat{\eta}\hat{\Delta}|E_{\chi,\vec{k}}^{(0)}\rangle.
\end{equation}

As discussed in the main text, the evolution of the excitations spectrum at a given $k$ point can be tied to the sign of the determinant of $\hat{V}_{\vec{k}}$.  The determinant can be expressed in a simple form by employing the states in Eqs. (\ref{eq:zero_states}) and considering that $\hat{V}$ is block-diagonal in the eigenbasis of
${\cal \hat{\chi}}_{\vec{k}}$, because $[\hat{V},{\cal \hat{\chi}}_{\vec{k}}]=0$. Hence, we have that
\[
\det\hat{V}_{\vec{k}}=\langle u_{L,\vec{k}}^{(0)}|\hat{\Pi}_{+}(\hat{\eta}\hat{\Delta})\hat{\Pi}_{+}|u_{L,\vec{k}}^{(0)}\rangle\langle u_{R,\vec{k}}^{(0)}|\hat{\Pi}_{-}(\hat{\eta}\hat{\Delta})\hat{\Pi}_{-}|u_{R,\vec{k}}^{(0)}\rangle,
\]
where $\hat{\Pi}_{\pm}=\frac{1}{2}(1-{\cal \hat{\chi}}_{\vec{k}})$
is a projector on $\chi=\pm1$ eigensubspace of ${\cal \hat{\chi}}_{\vec{k}}$.

\subsection{N\'eel spin pattern}

\begin{figure}[t!]
\includegraphics[clip,width=0.95\columnwidth]{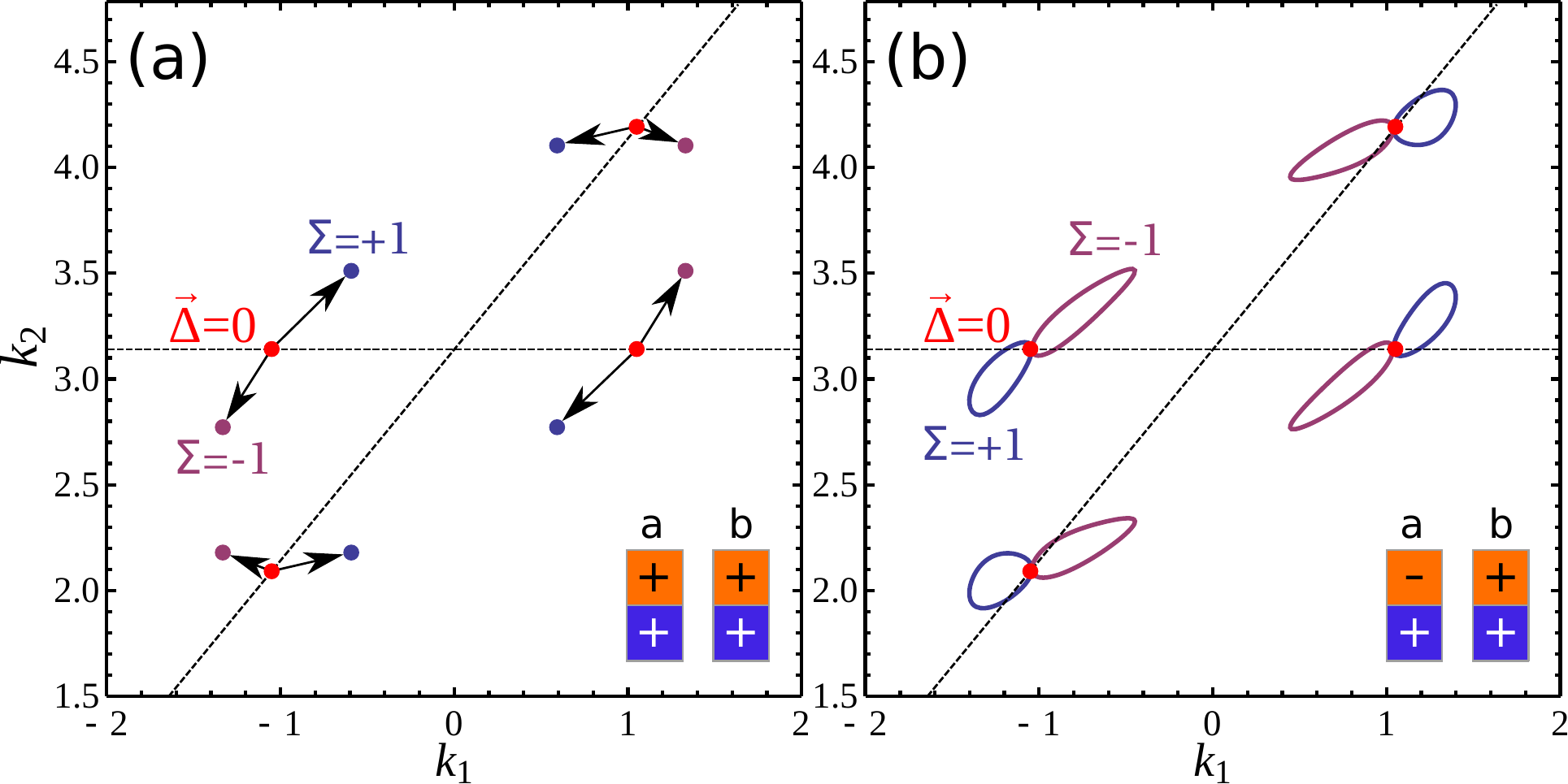}
\protect\caption{Representative nodal superconductivity evolving from a N\'eel AFM semimetal (see Fig. 1 in the main text) with Dirac points indicated by red dots. 
at zero 
(a) uniform case and (b) non-uniform configuration for the pairing amplitude in the unit cell. Site labels and schematic of the unit cell are in the center of the panels (blue and orange are for spin up and down). $a$ and $b$ refer to the orbitals and ($+$,$-$) indicate the sign of the pairing amplitude. 
Different $\Sigma$ sectors yield inequivalent point nodes or nodal rings in the superconducting state.
In case (a) both chiral symmetry operators commute with each other and with $\Sigma$,
so each $\Sigma$ sector is chiral symmetric at any $\hat{\Delta}$. In case
(b) both chiral symmetries are broken by $\hat{\Delta}$ and only their
product, i.e., the $\Sigma$ symmetry is preserved.\label{fig:FSaf}}
\end{figure}

In Fig. \ref{fig:FSaf} we report the excitations spectra of the coexistence state of N\'eel AFM state and s-wave pairing.
The Dirac AFM semimetal with point nodes is converted into a nodal superconductor with point nodes in each $\Sigma$ symmetric block (Fig. \ref{fig:FSaf}(a)) for a uniform pairing amplitude or into multiple nodal rings superconductor in the presence of orbitally inequivalent pairing amplitudes in the unit cell (Fig. \ref{fig:FSaf}(b)).
\\


\begin{thebibliography}{99}
\bibitem{footnote-so} The kinetic part of the Hamiltonian can include spin-orbital coupling breaking the SU(2) symmetry down to U(1) spin symmetry thus preserving axial spin symmetry, i.e. involving only one spin orientation. Such type of interactions may arise from the intra-atomic spin-orbit coupling in the presence of quenching of spin components due to lattice distortions and magnetic anisotropy or as lowest order mirror symmetric spin-orbit couplings as well as in the presence of confining symmetry breaking terms that couple orbital degrees of freedom to only one electron spin orientation.
\bibitem{SupMat} In the Appendix we provide details for the derivation of the criterion for nodal superconductors. For $z2$ spin pattern we explicitly report the matrix structure of $\hat{\eta}$ and $\hat{\chi}$ that build up the chiral symmetry operators. Then, a representative case of the evolution of the superconducting state for the N\'eel AFM state is also reported.
\end{thebibliography}
\end{document}